\documentclass[12pt,epsf]{article}

\usepackage[dvips]{graphicx}
\usepackage{color}

\usepackage[dvips]{graphicx}
\usepackage{color}

\newcommand{\beq}{\begin{equation}}
\newcommand{\eeq}{\end{equation}}
\newcommand{\bea}{\begin{eqnarray}}
\newcommand{\eea}{\end{eqnarray}}
\newcommand{\AmSLaTeXe}{%
 $\mathcal A$\lower.4ex\hbox{$\!\mathcal M\!$}$\mathcal S$-\LaTeXe}

\usepackage{bm}

\begin{document}
\thispagestyle{empty}
\vspace*{-15mm}
{\bf OCHA-PP-347}\\

\vspace{15mm}
\begin{center}
{\Large\bf
Vacuum Magnetic Birefringence Experiment as a probe of the Dark Sector\\
}
\vspace{7mm}

\baselineskip 18pt
{\bf Xing Fan${}^{1, 2}$, Shusei Kamioka${}^{1}$, Kimiko Yamashita$^{3,4}$, \\
Shoji Asai${}^{1}$, and Akio Sugamoto${}^{3, 5}$
}
\vspace{2mm}

{\it
${}^{1}$Department of Physics, Graduate School of Science, the University of Tokyo, 7-3-1 Hongo, Bunkyo-ku, Tokyo 113-0033, Japan\\
${}^{2}$Department of Physics, Harvard University, Cambridge, Massachusetts 02138, USA\\
${}^{3}$Department of Physics, Graduate School of Humanities and Sciences, Ochanomizu University, 2-1-1 Ohtsuka, Bunkyo-ku, Tokyo 112-8610, Japan\\
${}^{4}$Program for Leading Graduate Schools, Ochanomizu University, 2-1-1 Ohtsuka, Bunkyo-ku, Tokyo 112-8610, Japan\\
${}^{5}$Tokyo Bunkyo SC, the Open University of Japan, Tokyo 112-0012, Japan \\
}

\vspace{10mm}
\end{center}
\begin{center}
\begin{minipage}{14cm}
\baselineskip 16pt
\noindent
\begin{abstract}

Vacuum magnetic birefringence (VMB) is a nonlinear electromagnetic effect predicted by QED.
In addition to seeing the effect from QED, it is also possible for the measurement to probe the dark sector.
VMB as predicted by QED is parity conservative, but the effect from the dark sector can be parity violative.
To pursue this possibility, we calculated the effect from the dark sector with the generalized Heisenberg-Euler effective Lagrangian that is applicable to parity-violating theories in the weak and homogeneous field limit.

The contribution of the dark sector neutrinos in a dark sector model to the VMB experiment is studied.
The contribution comes from the mixing of the photon with the dark sector $Z$ boson and induces a parity-violating electromagnetic interaction.

Polarization change of a laser beam in an external magnetic field is studied, where the angle between the initial polarization and the magnetic field is 45$^{\circ}$ or $-45^{\circ}$.
The contribution from the dark sector modifies the magnitude of the polarization change, so it can be detected by measuring the magnitude precisely.

In addition to the change in the magnitude, the dark sector also induces parity-violating effects.
We also propose a new scheme to measure the effect of parity violation directly.
By measuring the change of polarization of a laser polarized in parallel or perpendicular to the applied magnetic field with a ring Fabry-P\'erot resonator, one can search for the effect directly.
If a signal appears in this scheme, it would be evidence of parity violation from beyond standard model theories.

\end{abstract}

\end{minipage}
\end{center}

\baselineskip 18pt
\def\thefootnote{\fnsymbol{footnote}}
\setcounter{footnote}{0}

\newpage

\section{Introduction}
So far, experimental results have agreed well with the standard model (SM), which is based on a $SU(3)_C \times SU(2)_L \times U(1)_Y$ gauge theory.
Unfortunately, even with the LHC operating at 13 TeV, no clear evidence of physics beyond the standard model has emerged.
There are, however, a number of phenomena which do not appear to arise from the SM.
Examples include the mass stability of the Higgs particle, mixing of quarks and leptons (including neutrinos), baryon asymmetry, dark matter, and dark energy.
These give ample motivation for searches for physics beyond the SM.

In this paper, we focus on possible dark sectors (DS), collections of fields whose couplings with SM particles are extremely weak, and whose corresponding particles represents dark matter.
To probe such a sector with ordinary SM matter, we must use observable processes which receive corrections from diagrams involving virtual DS particles.
Vacuum magnetic birefringence (VMB) is a good candidate process as it receives no tree-level contributions from QED.
High precision is needed to probe the higher order process with virtual particles.
We study the contribution from virtual DS fermion pairs in this process using the Heisenberg-Euler effective Lagrangian.
Two conditions should be satisfied to apply this: One is that the rate at which the field varies times Planck constant $h$ should be smaller than the energy of the mass of the lightest particle considered in the loop, 
\begin{equation}
h\nu_{\mathrm{laser}} \ll mc^2.
\label{lowenergycondition}
\end{equation}
The other one is that the coupling of the applied field with the virtual particle considered should be weaker than the square of the the mass of the lightest particle.
\begin{equation}
\hbar\epsilon  e|\bm{B}|\ll m^2c^2, 
\label{lowenergycondition2}
\end{equation}
where $\epsilon e$ is a coupling constant of the field and the particle. 
In this paper, we consider interactions of a photon about $h\nu_{\mathrm{laser}}= 1$ eV with an external magnetic field about $|\bm{B}|= 10$ Tesla.
Both of the conditions above are satisfied in QED and in the dark sector theory as we will outline in detail in Sec. 3 and 4.
Several experiments (BMV \cite{BMV}, PVLAS \cite{PVLAS}, and OVAL \cite{OVAL}) aim to measure VMB by observing this change in a laser's ellipticity.
Even though we assume the magnetic field is perfectly homogeneous in this paper, our discussions are applicable to the actual experiments, as laboratory magnetic fields are perfectly approximated as constant on length scales of the Compton wavelength $\lambda_{\rm C}=h/(mc)$.
These experiments intend to see the $O(\alpha^2)$ corrections ($\alpha$ is the fine structure constant in QED), which exist in the effective action of the electromagnetic field, by observing the change of polarization of the laser beam in a strong magnetic field.
The lowest order contribution from QED contains loops in the diagram and is quite small, while that from the DS could be large.
Although the DS gauge boson is produced virtually in VMB process, the sensitivity for the DS does not depend on the mass of the DS photon. 
\footnote{See Sec. 3 for the details of the calculations. In our scheme, an ordinary photon $\tilde{A}_{\mu}$ is a superposition of the DS gauge boson $B'_{\mu}$ and a $U(1)_{Y}$ gauge boson $B_{\mu}$. The $\tilde{A}_\mu$ converts to $B'_{\nu}$ and then couple to the DS fermion loop. It can be shown that the dependency on mass disappears by calculating the propagator $<\tilde{A}_{\nu}B'_{\nu}>$ properly.}

``Birefringence'' occurs when different polarizations of light have different refractive indices.
For example, let $n_{\parallel}$ be the refractive index for the polarization vector $\bm{\epsilon}_{\parallel}$ which is parallel to the external magnetic field $\bm{B}$ and $n_{\bot}$ be the other refractive index for the polarization vector $\bm{\epsilon}_{\bot}$ which is perpendicular to the field.
Since the refractive index $n$ and the phase velocity $v$ of light are inversely related as $n\cdot v=1(=c)$ (we use natural units hereafter), the difference of the refractive indices induces a phase shift between the two polarizations and changes the polarization of the light.
QED predicts birefringence in the presence of a magnetic field,
\begin{eqnarray}
n_{\parallel}-n_{\bot}=\frac{\alpha}{30\pi} \left(\frac{eB}{m_e^2}\right)^2 = 4 \times 10^{-24} (B~[T])^2,  \label{difference of refraction}
\end{eqnarray}
where $m_e$ is the mass of an electron and the subscripts refer to the angle between the polarization and the magnetic field. (Note that we neglect the small contributions from particles heavier than the electron)\cite{Baier1, Baier2, Nuovo, Toll}.
Two polarizations of light, $\bm{\epsilon}_{\parallel}$ and $\bm{\epsilon}_{\bot}$, at wavelength $\lambda$ will therefore pick up a phase difference $\Psi$ of
\begin{eqnarray}
\Psi = \pi(n_{\parallel}-n_{\bot}) \frac{L}{\lambda},
\end{eqnarray}
after traveling a distance $L$ through the field.

The phase shift $\Psi$ is of order $\alpha^2$, as can be seen in Eq. (\ref{difference of refraction}) above. 
Assuming $B=1$ Tesla, the magnitude of birefringence is $\Delta n = 4.0\times 10^{-24}$, which is as small as the distortion from gravitational waves. 
Thought the effect is tiny and has not yet been observed, recently development of high finesse (more than $10^5$) Fabry-P\'erot resonators as well as strong magnetic fields of 10 Tesla significantly enhanced the sensitivity of VMB searches.
The VMB experiments are now at the stage where they would need a improvement in sensitivity of 20-500 to observe VMB.

Note that light-shining-through-a-wall experiments are similar types of experiments to VMB and have sensitivity on minicharged particles that couple to photon when the mass of the minicharged particle is less than the energy of the photon\cite{lsw1, lsw2}.
In our case, we consider the exclusion limit to particles that have larger mass than the energy of the probe photon.
LSW experiments and VMB experiments are complementary in search of new particles.

The formula in Eq. (\ref{difference of refraction}) can be derived using the QED effective Lagrangian $\mathcal{L}^{\mathrm{QED}}_{\mathrm{eff}}$ found in 1936 by W. Heisenberg and H. Euler\cite{Heisenberg-Euler, review of H-E}:
\begin{eqnarray}
\mathcal{L}^{\mathrm{QED}}_{\mathrm{eff}}=-\mathcal{F} +\frac{8}{45}\left(\frac{\alpha^2}{m_e^4}\right)\mathcal{F}^2 + \frac{14}{45}\left(\frac{\alpha^2}{m_e^4}\right)\mathcal{G}^2+ \cdots.  \label{H-E}
\end{eqnarray}
Here, 
\begin{eqnarray}
\mathcal{F} =\frac{1}{4}F_{\mu\nu}F^{\mu\nu}=\frac{1}{2}\left(\bm{B}^2-\bm{E}^2\right), ~\mathcal{G} =\frac{1}{4}F_{\mu\nu} \tilde{F}^{\mu\nu}=\bm{E} \cdot \bm{B},  \label{F and G}
\end{eqnarray}
where the dual field strength is defined by $\tilde{F}_{\mu\nu}\equiv \frac{1}{2}\epsilon_{\mu\nu\lambda\rho}F^{\lambda\rho}$, and $\epsilon_{\mu\nu\lambda\rho}$ is a totally anti-symmetric tensor with $\epsilon_{0123}=1$.
The first term in Eq. (\ref{H-E}) is the usual kinetic contribution from our gauge fields.
The second term and third term induce VMB (See also \cite{Dittrich, Dunne, Rizzo1, Karbstein}).

We know that parity is a symmetry of QED, but a DS might not have this symmetry.
Consequently, Eq. (\ref{H-E}) needs to be generalized to include parity-violating terms. 
We do this below, following another work, ``Generalized Heisenberg-Euler (H-E) formula'' \cite{generalized H-E}.  

In this paper we establish a DS model in which the DS contribution to VMB is of the same kind as that of QED.
In order to exemplify our method for probing the DS, we introduce a scalar field having both hypercharges of the SM and that of the DS.
As a result, the mixing between the real photon and the $U(1)$ hypercharge gauge boson $B'_{\mu}$ in the DS is introduced.
We also explore the interesting case in which the DS contribution to VMB differs qualitatively from that of QED, making the two separable.
We study such a model, apply the generalized Heisenberg-Euler formula to it, and study how the magnetic birefringence effect behaves.
In the course of this study, we propose a new experiment using a "ring resonator", which is more effective than the conventional Fabry-P\'erot resonator to detect parity-violating effects.

\section{Generalized Heisenberg-Euler formula} 
In this section, we review the generalized Heisenberg-Euler (H-E) formula for a model in which a fermion field $\psi(x)$ couples to a gauge field $A_{\mu}(x)$ in a general way, with a vector coupling $g_V$ and an axial vector coupling $g_A$.  
This formula is applicable to parity violating models in general.
The action for this model is
\begin{eqnarray}
S_{\psi}(m)=\int d^4 x~ \bar{\psi} \left[\gamma^{\mu} \left( i\partial_{\mu} -(g_V+g_A \gamma_5) A_{\mu}\right) - m \right] \psi(x), \label{original action}
\end{eqnarray}
which gives an effective action $S_{\mathrm{eff}}[A_{\mu}]$ and an effective Lagrangian $\mathcal{L}_{\mathrm{eff}}[A_{\mu}]$ for background field configuration $A_{\mu}(x)$ as 
\begin{eqnarray}
 S_{\mathrm{eff}}[A_{\mu}] &=&\int d^4x~\mathcal{L}_{\mathrm{eff}}[A_{\mu}]= -i \ln \left[ \int \mathcal{D} \psi(x) \mathcal{D} \bar{\psi}(x) e^{iS_{\psi}(m)} \right] \nonumber \\
&=&-i \mathrm{Tr} \ln \left[\gamma^{\mu} \left( i\partial_{\mu} -(g_V+g_A \gamma_5) A_{\mu}\right) - m \right].
 \end{eqnarray}

Following faithfully the derivation by J. Schwinger \cite{Schwinger} of the effective action in the proper-time formalism \cite{Fock and Nambu}, with a small simplification in terms of the path integral \cite{Feynman}, the following results are obtained in \cite{generalized H-E}, which gives the leading order contribution in the perturbative limit, corresponding to
the effective interaction of four $A$ fields:

\begin{eqnarray}
\mathcal{L}^{(2)}_{\mathrm{eff, ~general}}=a ~\mathcal{F}^2 +b ~\mathcal{G}^2
+i c ~\mathcal{F}\mathcal{G},  \label{generalized H-E}
\end{eqnarray}
with
\begin{eqnarray}
&&a=\frac{1}{(4\pi)^2m^4}\left(\frac{8}{45}~ g_V^4- \frac{4}{5}~g_V^2g_A^2 - \frac{1}{45}~g_A^4\right), \label{a} \\
&&b=\frac{1}{(4\pi)^2m^4}\left (\frac{14}{45} ~g_V^4 + \frac{34}{15}~ g_V^2 g_A^2 + \frac{97}{90}~g_A^4 \right), \label{b} \\
&&c=\frac{1}{(4\pi)^2 m^4} \left(\frac{4}{3}~g_V^3g_A + \frac{28}{9} ~g_V g_A^3 \right), \label{c}
\end{eqnarray}
The original H-E formula Eq. (\ref{H-E}) is reproduced with $g_V=-e$ and $g_A=0$. 
In case of V+A or V-A interactions, we set $g_A= \pm g_V$, respectively, and we have
\begin{eqnarray}
&&a(V \pm A)=-\frac{29}{45}\left(\frac{g_V^2}{4\pi m^2}\right)^2,~ b(V \pm A)=\frac{329}{90}\left(\frac{g_V^2}{4\pi m^2}\right)^2, ~\mathrm{and}\nonumber \\
&&c(V \pm A)=\pm \frac{40}{9}\left(\frac{g_V^2}{4\pi m^2}\right)^2. \label{V+-A}
\end{eqnarray}

Here, $\mathcal{F}^2$ and $\mathcal{G}^2$ are symmetric under a parity transformation (transformation with respect to the space inversion, $\bm{x} \to -\bm{x}$) while $\mathcal{F}\mathcal{G}$ is not.
The coefficients $a$ and $b$ describe the usual parity conserving effects, while $c$ gives the parity violating component.
Also, note that the contribution of $\mathcal{F}\mathcal{G}$ in the effective action is purely imaginary.  
Using the above effective action with the constants $(a, b, c)$, we can estimate the refractive indices $n_{\parallel}$ and $n_{\bot}$ (see Section 4).  

\section{Dark Sector Model}
The DS (dark sector) is a sector which interacts weakly with SM particles.
Experimental restrictions on the DS are not strong, so various models can be constructed. 
In order to achieve renormalizability in the DS, the DS theory needs to be anomaly-free.

We use Eq. (\ref{original action}) as the DS action.
Hereafter, we add primes to DS fields, coupling constants, and masses (e.g. $A'_{\mu}(x)$, $\psi'(x)$, $g'$, and $m'$).
Rewriting the DS Lagrangian according to this convention, we have
\begin{eqnarray}
\mathcal{L}_{\psi'}=\bar{\psi}' \left[\gamma^{\mu} \left( i\partial_{\mu} -(g'_V+g'_A \gamma_5) A'_{\mu}\right) - m' \right] \psi'(x), \label{fermionic Lagrangian}
\end{eqnarray}

An anomaly arises from a three point function of the gauge fields.
A convenient way to discuss this is to decompose the fermion field into right-handed one $\psi_R$ and left handed one $\psi_L$.
They are defined as
\begin{equation}
\psi_R \equiv \frac{1+\gamma^5}{2}\psi, ~~\psi_L \equiv \frac{1-\gamma^5}{2}\psi.
\end{equation}
In this expression, the Lagrangian above is written as,
\begin{equation}
\mathcal{L}_{\psi'}=\bar{\psi}'_R \gamma^{\mu}\left( i\partial_{\mu}-g'_1Y'_R A'_{\mu} \right)  \psi'_R + \bar{\psi}'_L\gamma^{\mu}\left( i\partial_{\mu}-g'_1Y'_L A'_{\mu} \right) \psi'_L + m \left(\bar{\psi}'_R\psi'_L+\bar{\psi}'_L\psi'_R\right).
\end{equation}
where $g'_{1}$ is the coupling constant of the fermions and the gauge field, and  $Y'_{R_i}$ and $Y'_{L_i}$ are the charges of R- and L-handed fermions respectively, defined by
\begin{eqnarray}
\frac{1}{2}(g'^i_{V}+g'^i_{A})=g' _1Y'_{R_i}, ~~\frac{1}{2}(g'^i_{V}-g'^i_{A})=g'_1 Y'_{L_i}.
\end{eqnarray}.

Right-handed and left-handed fermions contribute to this anomaly with opposite signs.
Therefore, the anomaly cancellation condition is
\begin{eqnarray}
\sum_{R_i} (Y'_{R_i})^3-\sum_{L_i} (Y'_{L_i})^3=0,
\end{eqnarray}
where ``$i$" is introduced to represent different species of fermions in general.
We can generate a variety of anomaly-free models in which gauge fields couple $A'_{\mu}(x)$ to the DS in a general way.

The magnitude of VMB could be affected by the presence of a dark sector when $A'$ couples to the photon.
While kinetic mixing is usually used \cite{Holdom}, we proceed here with the simpler mass mixing since the discussion can be done at tree level with this approach. (A similar discussion is possible using the kinetic mixing obtained from one-loop corrections. See the footnote.)

Here we introduce a scalar field, $S(x)$, which couples to the SM $U(1)_Y$ gauge field $B_{\mu}(x)$ as well as to the DS $U(1)'_{Y'}$ gauge field $B'_{\mu}(x)$ with charges $g_1Y_s$ and $g'_1 Y'_s$ respectively:
\begin{eqnarray}
\mathcal{L}_S=\left\vert \left(i \partial_{\mu} -g_1 Y_s B_{\mu} - g_1^{'}Y'_s B'_{\mu} \right) S(x) \right\vert^2.
\end{eqnarray}
If $S$ takes a vacuum expectation value $\langle S \rangle= v_s/\sqrt{2}$, then the gauge fields $B_\mu(x)$ and $B'_\mu(x)$ mix as follows:
\begin{eqnarray}
\mathcal{L}_{\mathrm{mixing}}= v_s^2 \left\{ \frac{1}{2}(g_1Y_s)^2 B_{\mu} B^{\mu} + (g_1 Y_s) (g'_1 Y'_s) B_{\mu} B^{' \mu} + \frac{1}{2}(g'_1Y'_s)^2 B'_{\mu} B^{' \mu}  \right\}.
\end{eqnarray}
Since the DS should not change the parameters in the SM so much, we assume the mixing parameter $\varepsilon$, defined in the following, is extremely small, 
\begin{eqnarray}
\varepsilon \equiv \frac{g_1 Y_s}{g'_1Y'_s} \ll 1.
\end{eqnarray}
Then we have
\begin{eqnarray}
\mathcal{L}_{\mathrm{mixing}}=\frac{1}{2}\; m_{B '}^2 \left( \varepsilon^2B_{\mu} B^{\mu} + 2\varepsilon B_{\mu} B^{' \mu} + B'_{\mu} B^{' \mu} \right), \label{mixing Lagrangian}
\end{eqnarray}
with $m_{B '}= g'_1 Y'_s v_s$.\footnote{If the mixing Lagrangian is obtained by kinetic mixing as $\mathcal{L}'_{\mathrm{mixing}}=-\frac{1}{4} A_{\mu\nu}(x)A^{\mu\nu}(x)+\varepsilon A_{\mu\nu}(x)A^{'\mu\nu}(x)-\frac{1}{4} A'_{\mu\nu}(x)A^{'\mu\nu}(x) + \frac{1}{2} m_{B'}^2 A'_{\mu}(x) A^{'\mu}(x)$, it becomes $\mathcal{L}''_{\mathrm{mixing}}=-\frac{1}{4} B_{\mu\nu}(x)B^{\mu\nu}(x)-\frac{1}{4} B'_{\mu\nu}(x)B^{'\mu\nu}(x) + \frac{1}{2} m_{B'}^2 (\varepsilon B_\mu + B'_\mu)^2$ by the transformation $A_\mu=B_\mu+ \varepsilon B'_\mu$,  and $A'_\mu=B'_\mu+ \varepsilon B_\mu$, which is not orthogonal. This reproduces the mass mixing Lagrangian in Eq. (\ref{mixing Lagrangian}). }
To identify the SM photon, we have to diagonalize the $3\times3$ matrix including all terms in the Lagrangian quadratic in the fields $B_\mu(x)$, $B'_\mu(x)$, and $A^3_{\mu}(x)$.
Here, the field $B_\mu(x)$ denotes the gauge field of the $U(1)_Y$, and $A^3_{\mu}(x)$ denotes the third component of  $SU(2)_L$ in the SM theory.
Since we are considering the mixing of the DS boson $B'_\mu(x)$ with $B_\mu(x)$, we need to diagonalize the $3\times3$ matrix to see the effect of the mixing on the usual "photon".
The mixing matrix appears in the mass part of the Lagrangian,
\begin{eqnarray}
\mathcal{L}_{mass}=\frac{v^2}{8} \pmatrix{A^{3\mu}(x),\;B^\mu(x),\;B'^\mu(x)} \pmatrix{g_2^2&&-g_1g_2&&0 \cr -g_1g_2&&g_1^2+\alpha^\prime \varepsilon^2&&\alpha^\prime \varepsilon \cr 0&&\alpha^\prime\varepsilon&&\alpha^\prime } \pmatrix{A^3_\mu(x) \cr B_\mu(x) \cr B'_\mu(x)},
\end{eqnarray}
where $\alpha^\prime=4(m_{B'}/v)^2$ and $v$ is the vacuum expectation value of the SM Higgs.
The mass eigenstates are denoted with a tilde as $(\tilde{A}, \tilde{Z}, \tilde{B}')$, and the corresponding masses squared are
\begin{eqnarray}
&&(m_{\tilde{A}})^2=0,~ (m_{\tilde{Z}})^2= \frac{1}{4} v^2 (g_1^2 +g_2^2) + \varepsilon^2 \frac{g_1^2}{g_1^2+g_2^2-\alpha^\prime} (m_{B'})^2, ~\mathrm{and}\\
&&(m_{\tilde{B'}})^2=(m_{B'})^2 \left(1+\varepsilon^2 \frac{g_2^2-\alpha^\prime}{g_1^2+g_2^2-\alpha^\prime} \right).
\end{eqnarray}
The $W^{\pm}$ fields are not modified by the introduction of the mixing, so $m_W^2=\frac{1}{4}v^2 g_2^2 $ as usual. 
Thus
\begin{eqnarray}
\frac{(m_W)^2}{(m_{\tilde{Z}})^2}=\frac{g_2^2}{g_1^2+g_2^2} \times \left\{ 1- \varepsilon^2 \frac{4g_1^2}{(g_1^2+g_2^2)(g_1^2+g_2^2-\alpha^\prime)} \left(\frac{m_{B'}}{v}\right)^2+{\cal O}(\epsilon^4) \right\}.\label{masswinberg}
\end{eqnarray}

The mass eigenstate $\tilde{A}_\mu$ can be expressed as
\begin{eqnarray}
\tilde{A}_\mu=\frac{g_1 A^3_\mu +g_2 B_\mu}{\sqrt{g_1^2+g_2^2}}-\varepsilon \frac{g_2}{\sqrt{g_1^2+g_2^2}}B'_\mu
\end{eqnarray}

The real photon is the diagonalized $\tilde{A}_\mu$, and this is the modified SM photon.
The modified photon can convert to the DS gauge boson $B'_\mu$, with a mixing parameter $\chi$
\begin{eqnarray}
\chi=-\varepsilon \cos \theta_W,
\end{eqnarray}
where $\cos \theta_W=g_2/\sqrt{g_1^2+g_2^2}$.
This mixing parameter comes in each conversion from real photon to the DS $U(1)'_{Y'}$ gauge boson $B'_\mu$. 

The effective action of DS fermions coupled to the modified SM photon can be written as \footnote{Consider a Feynman diagram in which four real photons ($A^\mu$) are
coming in. Each photon converts to the DS boson ($B'^\mu$) with the mixing parameter $\chi$. These four DS bosons couple to a loop of the DS fermions $\psi'$ and induce the terms in the bracket of Eq. (29).}
\begin{eqnarray}
\mathcal{L'}_{\mathrm{eff}}=\chi^4 \left\{a ~\mathcal{F}^2 +b ~\mathcal{G}^2
+i c ~\mathcal{F}\mathcal{G} \right\}. \label{applied to DS}
\end{eqnarray} 
The coefficients $(a, b, c)$ are given in Eqs.(\ref{a})-(\ref{c}), where $m$, $g_V$ and $g_A$ are those of the DS fermions.
Note that the contribution of the lightest fermion to VMB is the largest, and thus we apply this calculation to the lightest fermion explicitly later.

The mixing parameter $\chi$ and the DS gauge boson mass $m_{\tilde{B'}}$ are restricted by various experiments\cite{HP 1}\cite{HP 2}. 
Figures 1 and 2 in \cite{HP 1} show that the parameter space not excluded by the experiments they consider is as follows:
\begin{eqnarray}
\chi \le 10^{-6}~\mathrm{for}~ m_{\tilde{B'}} \ge 1 ~\mathrm{MeV} , ~\mathrm{or} \\
10^{-6} \le \chi \le 10^{-3} ~\mathrm{for}~ m_{\tilde{B'}} \ge 1~ \mathrm{GeV}. \label{experimental constraints}
\end{eqnarray}

Other constraints come from the modification of SM parameters, which arises from the mixing with the real world and the DS. 
To do this we have to know the mass eigen-state of $\tilde{Z}$, which is 
\begin{eqnarray}
\tilde{Z}_\mu=\frac{g_2 A^3_\mu-g_1B_\mu}{\sqrt{g_1^2 + g_2^2}}-\varepsilon \frac{g_1 \alpha^\prime}{(g_1^2 +g_2^2 -\alpha^\prime)\sqrt{g_1^2 + g_2^2}} B'_\mu.
\end{eqnarray}  

The DS modifies the value of Weinberg angle $\theta_W$, see also Eq. (\ref{masswinberg}).
The Weinberg angle in our model $\tilde{\theta}_W$ is
\begin{eqnarray}
\cos^2\tilde{\theta}_W&=&\frac{(m_W)^2}{(m_{\tilde{Z}})^2}
\end{eqnarray}
This should be consistent, to within the experimental error of $O(10^{-4})$ (on-shell scheme), with the SM value.
By using Eq. (\ref{masswinberg}) and the typical values from standard model, $v\simeq 246$~GeV, $g_1\simeq 0.344$, and $g_2\simeq 0.641$, we get the limit on the mixing parameter $\chi$ as
\begin{eqnarray}
\chi m_{B'} < 2 ~\mathrm{GeV}
\end{eqnarray}

Thus, the constraint from the SM is not strong.
If we take $m_{B'}= 1 ~\mathrm{GeV}$, even the maximum value of the mixing parameter $\chi=10^{-3}$ can satisfy the constraints on the weak interaction of the SM.
The constraints from Eq. (\ref{experimental constraints}) are much stronger than that from the Weinberg angle.

The fermion mass term violates the gauge symmetry 
\begin{eqnarray}
&&\psi(x) \to \psi'(x)= e^{i (g_V + \gamma_5 g_A) \theta (x) }, ~\mathrm{or~equivalently} \\
&&\psi'_R(x)=e^{i g'_1Y'_R \theta (x) }\psi_R(x),~\psi'_L(x)=e^{i g'_1Y'_L \theta (x) }\psi_L(x), ~\mathrm{and} \\
&&B'_{\mu}(x) \to (B')'_{\mu}(x) =  B'_{\mu}(x) - \partial_{\mu} \theta (x).
\end{eqnarray}
Therefore, the fermion mass should be generated via the spontaneously breakdown of the gauge symmetry. 
We introduce a ``Higgs field" $S(x)$ and consider the gauge invariant Lagrangian for a fermion with Yukawa coupling $y$ to the Higgs field,
\begin{eqnarray}
\mathcal{L}_{\psi}&=&\bar{\psi}_R \gamma^{\mu} \left( i \partial_{\mu} - g'_1 Y'_R (B')'_{\mu} \right) \psi_R + \bar{\psi}_L \gamma^{\mu} \left( i \partial_{\mu} - g'_1 Y'_L (B')'_{\mu} \right) \psi_L  \nonumber \\
&+& y \bar{\psi}_R S \psi_L +  y \bar{\psi}_L S^{\dagger} \psi_R+|(i\partial_\mu -g'_1Y'_S (B')'_\mu) S |^2-V_S(S^{\dagger}S).
\end{eqnarray}
Assume the Higgs field $S$ has charge $Y'_{S}=Y'_R-Y'_L$. $\mathcal{L}_{S}$ is gauge invariant and the fermion mass $m$ is generated spontaneously by the vacuum expectation value of $\phi$ as $m=y\langle S \rangle=y v_{S}$.

Therefore, the generalized H-E formula can be applied to the various models of DS with the help of the spontaneous breaking, and it can handle the components of the full effective action of the models.  
The main contribution to the effective action comes from a fermion with the smallest mass in the DS, which might be neutrino-like fermions.
At this point we surely recognize that it is an interesting possibility for the DS to mimic the SM: the DS has the same gauge group $SU(3)' _{C'} \times SU(2)'_{L'} \times U(1)'_{Y'}$ and the same field content, but the masses $m'$ and couplings $g'_3$, $g'_2$, and $g'_1$ differ from the SM. Many people have considered this possibility\cite{RFootRVolkas, RFoot}.  See a comprehensive review on DS \cite{HP 2}.

This model is manifestly anomaly free.
Here, we assume that the lightest fermion is the dark electron-type neutrino ``$\nu'_e$'', and no neutrino mixing appears in the DS for simplicity.
Then the lightest dark particle (LDP) is $\nu'_e$'.
It couples to the $Z'$ gauge boson in the DS with a V-A coupling, namely
\begin{eqnarray}
\mathcal{L}_{\nu'}&=& \bar{\nu'}_L \gamma^{\mu} \left( i \partial_{\mu} - \sqrt{(g'_1)^2 + (g'_2)^2} \frac{1}{2} Z'_{\mu} \right) \nu'_L \nonumber \\
&+& \bar{N'}_R  \gamma^{\mu} \left(i\partial_{\mu}\right) N'_R - m_{\nu} \bar{N'}_R \nu'_L,
\end{eqnarray} 
where the right-handed neutrino field $N'_R$ is introduced and the neutrino mass is considered to be Dirac.
If the DS is like this, then the VMB experiment is quite interesting.
The DS neutrino mass $m_{\nu'}$ can be much lighter than the electron in QED, and its coupling to $Z'$ is V-A in nature and violates parity symmetry.
The generalized H-E formula works well and the obtained parity-violating result differs from that of parity-conserving QED.
In this case, the effective Lagrangian, including the mixing between $Z'$ and $B'$, is
\begin{eqnarray}
\mathcal{L}_{\mathrm{eff}}(\nu'_e)=\chi_{\nu'}^4 \{a(V-A) \mathcal{F}^2 +b(V-A) \mathcal{G}^2 +i c(V-A)\mathcal{F}\mathcal{G} \},  \label{applied to DS neutrino}
\end{eqnarray}
where $\chi_{\nu'}=-\chi \sin \theta'_W=\varepsilon \cos \theta_W \sin \theta'_W$, and $a(V-A)$, $b(V-A)$, $c(V-A)$ are given in Eq. (\ref{V+-A}).

Roughly speaking, all the coupling constants are $O(1)$, so that the contribution from the DS electron-type neutrino might dominate over the QED contribution.
For instance, if
\begin{eqnarray}
\frac{m_{\nu'_e}}{m_e} \le \chi \le 10^{-3} - 10^{-6},
\end{eqnarray}
the smallness of the mixing parameter can be compensated for by the smallness of the DS neutrino mass.
Hence the DS neutrino with a mass of eV or keV could be an interesting target for the vacuum magnetic birefringence experiment.

In addition, as for the mass of the dark photon $m_{B'}$, since the mass does not appear in the observables, the parameter region where $m_{B'} \approx 1~\mathrm{GeV}$ and $\chi < 10^{-3}-10^{-4}$ is possible.
This is the region of the unified DM \cite{universal DM}, which can explain a number of astrophysical experiments in a single theory.
Therefore, the VMB experiment is sensitive to the scenario where the unified DM model has an LDP with 10-1000 eV and the heavier one with $\approx 800 \mathrm{GeV}$, via the see-saw like mechanism.


\section{Probe of Dark Sector via Vacuum Magnetic Birefringence experiments}

The effect from the DS Lagrangian also appears as changes in the refractive indices of the vacuum in the presence of an external magnetic field.
We consider the propagation of photons, with an energy of about 1 eV, in a magnetic field about 10 Tesla.
When an external magnetic field is applied perpendicularly to the propagation axis of the light, the eigenvectors of the polarization vector are $\bm{\epsilon}_{+}$ or $\bm{\epsilon}_{-}$, which are given in Appendix.
Their refractive indices are given by
\begin{eqnarray}
n_{\pm}&=&1+ \frac{1}{2} \bm{B}^2 \left\{ (a+b) \pm \sqrt{(a-b)^2-c^2} \right\}.\label{eq_refractiveindex}
\end{eqnarray}
See the Appendix for the detailed calculations.
The constants $(a, b, c)$ are sums of the contribution from QED and that from DS.
\begin{equation}
a=a_{\mathrm{QED}}+\chi^4a_{\mathrm{DS}\nu'},~~b=b_{\mathrm{QED}}+\chi^4b_{\mathrm{DS}\nu'},~~c=c_{\mathrm{QED}}+\chi^4c_{\mathrm{DS}\nu'},
\label{expressionsabc}
\end{equation}
 where the subscripts "QED" and "DS$\nu'$" represent the contribution from the two Lagrangian.
The contribution from QED is given by Eqs.(\ref{a})-(\ref{c}) with $g_V=-e$ and that from DS is given by the same equations by replacing the mass $m$ and coupling constants $g_A$ and $g_V$ by those in the DS Lagrangian and multiplying by $\chi^4$, see also Eq. (13).
We consider several DS neutrino masses where the condition Eq. (\ref{lowenergycondition}) and Eq. (\ref{lowenergycondition2}) hold.
Note that the coupling coefficient $\epsilon$ in Eq. (\ref{lowenergycondition2}) is $\chi$ in this case.

Whether the refraction constant $n$ is real or complex is determined by a discriminant $D=(a-b)^2-c^2$.
If $D > 0$, then $n$ is real. 
If $D<0$, $n$ has an imaginary part.
Denoting $x=g_A/g_V$, we get
\begin{eqnarray}
&&D=\left(\frac{g_V^2}{4\pi m^2}\right)^2 \frac{1}{45} f(x)f(-x), \\
&& f(x)=-6-138~x^2-99/2~x^4+60~x+ 140~x^3.
\end{eqnarray}
The region in which $D > 0$ can be numerically estimated as $|x| < 0.137935$ and $|x| > 1.43005$.
For QED, $x=0$ and $n$ is real, while for the V-A theory of ``$\nu_e'$'' neutrino , $x=-1$ and the refractive indices have imaginary parts.
In the SM mimic model for the DS, $x$ is given for the particle with the third component of isospin $I_3'$ and the charge $Q'$ as
\begin{eqnarray}
x=g_A/g_V=\frac{-I_3'}{I_3'- Q' \sin^2 \theta_W'}.
\end{eqnarray}

\subsection{Polarization change in a Vacuum Magnetic Birefringence configuration}
Here, we discuss the propagation of light in an external magnetic field $B$ in both cases: $D>0$ and $D<0$.
We consider the effect of magnetic field up to $O(|B|^2)$.
When $|D|<B^2\times O(a, b, c) \simeq 10^{-24}(B~[\mathrm{T}])^2$, the the difference of $n_{+}$ and $n_{-}$ is as small as $10^{-48}(B~[\mathrm{T}])^4$ and the space is almost isotropic.
Since this region is quite narrow, we mainly consider the case $|D| \gg B^2\times O(a, b, c)$ in the following discussions and refer to this region later.

In the following discussions, the polarization vector parallel to the magnetic field is denoted by $\bm{\epsilon}_{\parallel}$, and the one perpendicular to the magnetic field is denoted by $\bm{\epsilon}_{\bot}$. 
The eigenvectors  $\bm{\epsilon}_{\pm}$ can be calculated as
\begin{eqnarray}
\bm{\epsilon}_{\pm} \propto \left\{\begin{array}{ll}
-ic\bm{\epsilon}_{\parallel} + \left(a-b\pm\sqrt{(a-b)^2-c^2}\right)\bm{\epsilon}_{\bot}& (D>0)\\
-ic\bm{\epsilon}_{\parallel} + \left(a-b\pm i\sqrt{c^2-(a-b)^2}\right)\bm{\epsilon}_{\bot} & (D<0)
\end{array} \right.
\end{eqnarray}
The refractive indices are given by Eqs. (\ref{eq_refractiveindex}).

In a VMB experiment, the polarization of light is aligned to be at an angle of $45^\circ$ or $-45^\circ$ from the direction of the external magnetic field, $\bm{\epsilon}({45^{\circ}})\equiv \frac{1}{\sqrt{2}} (\bm{\epsilon}_{\parallel}+\bm{\epsilon}_{\bot})$. and $\bm{\epsilon}(-45^{\circ}) \equiv \frac{1}{\sqrt{2}} (\bm{\epsilon}_{\parallel}-\bm{\epsilon}_{\bot})$. 
The polarization vectors, after a distance $L$ though the magnetic field, can be written as
\begin{eqnarray}
\bm{\epsilon}(45^{\circ}) &\rightarrow& \left\{\begin{array}{ll}
\left(\cos(\Psi-2\phi)\bm{\epsilon}(45^{\circ}) -i\sin\Psi\bm{\epsilon}(-45^{\circ})\right)/\cos2\phi & (D>0)\\\label{eq_traditional1}
\left((\cosh\theta\sinh\Psi-\cosh\Psi)\bm{\epsilon}(45^{\circ})-i\sinh\theta\sinh\Psi\bm{\epsilon}(-45^{\circ}) \right)/\cosh\theta & (D<0)
\end{array} \right.\nonumber\\
&~&\\
\bm{\epsilon}(-45^{\circ}) &\rightarrow& \left\{\begin{array}{ll}
\left( -i\sin\Psi\bm{\epsilon}(45^{\circ}) +\cos(\Psi+2\phi)\bm{\epsilon}(-45^{\circ})\right)/\cos2\phi & (D>0)\\\label{eq_traditional2}
\left(i\sinh\theta\sinh\Psi\bm{\epsilon}(45^{\circ})+(\sinh\Psi+\cosh\theta\cosh\Psi)\bm{\epsilon}(-45^{\circ})\right)/\cosh\theta  & (D<0)
\end{array} \right.\nonumber\\
&~&
\end{eqnarray}
Here, $\Psi$ and $\phi$ are defined as 
\begin{eqnarray}
\sin \phi &=& \frac{c}{\left\{ \left( (a-b) + \sqrt{(a-b)^2 - c^2} \right)^2 +c^2 \right\}^{\frac{1}{2}}} \\
\cos \phi &=& \frac{(a-b) + \sqrt{(a-b)^2 - c^2} }{\left\{ \left( (a-b) + \sqrt{(a-b)^2 - c^2} \right)^2 +c^2 \right\}^{\frac{1}{2}}}\\
\Psi &=& \pi |B|^2 \frac{L}{\lambda} \sqrt{|(a-b)^2 -c^2|}, \label{ellipticity}
\end{eqnarray}
and $\theta$ is defined, only when $D<0$, as 
\begin{eqnarray}
\sinh\theta &=& \frac{a-b}{\left(c^2-(a-b)^2\right)^{\frac{1}{2}}}\times sgn\left(c\right)\\
\cosh\theta &=& \frac{|c|}{\left(c^2-(a-b)^2\right)^{\frac{1}{2}}}, \\
\end{eqnarray}
where $\lambda$ is the wavelength of the laser beam.
Note that the term $sgn\left(c\right)$ gives 1 when $c>0$ and $-1$ when $c<0$. 
After going through the magnetic field, the light acquires ellipticity
\begin{eqnarray}
\sin\Psi/\cos(\Psi-2\phi) &\textrm{for}& \bm{\epsilon}_i = \bm{\epsilon}(45^{\circ})~ (D>0)\label{eq_ellip1}\\
\sin\Psi/\cos(\Psi+2\phi) &\textrm{for}& \bm{\epsilon}_i = \bm{\epsilon}(-45^{\circ}) ~(D>0)\label{eq_ellip2}\\
\sinh\theta\sinh\Psi/(\cosh\Psi-\cosh\theta\sinh\Psi) &\textrm{for}& \bm{\epsilon}_i = \bm{\epsilon}(45^{\circ})~ (D<0)\label{eq_ellip3}\\
\sinh\theta\sinh\Psi/(\cosh\Psi+\cosh\theta\sinh\Psi) &\textrm{for}& \bm{\epsilon}_i = \bm{\epsilon}(-45^{\circ}) ~(D<0)\label{eq_ellip4}
\end{eqnarray}
Here, $\bm{\epsilon}_i$ is the initial injection polarization vector.
These reproduce the QED predicted VMB with $g_V=-e$ and $g_A=0$.

We define an ellipticity parameter $k_1$ as 
\begin{equation}
k_1= (ellipticity)\lambda/\pi|B|^2L,
\end{equation}
where the $(ellipticity)$ is given in one of Eqs. (\ref{eq_ellip1})-(\ref{eq_ellip4}).

\begin{figure}[]
\begin{center}
\resizebox{\textwidth}{!}{%
  \includegraphics{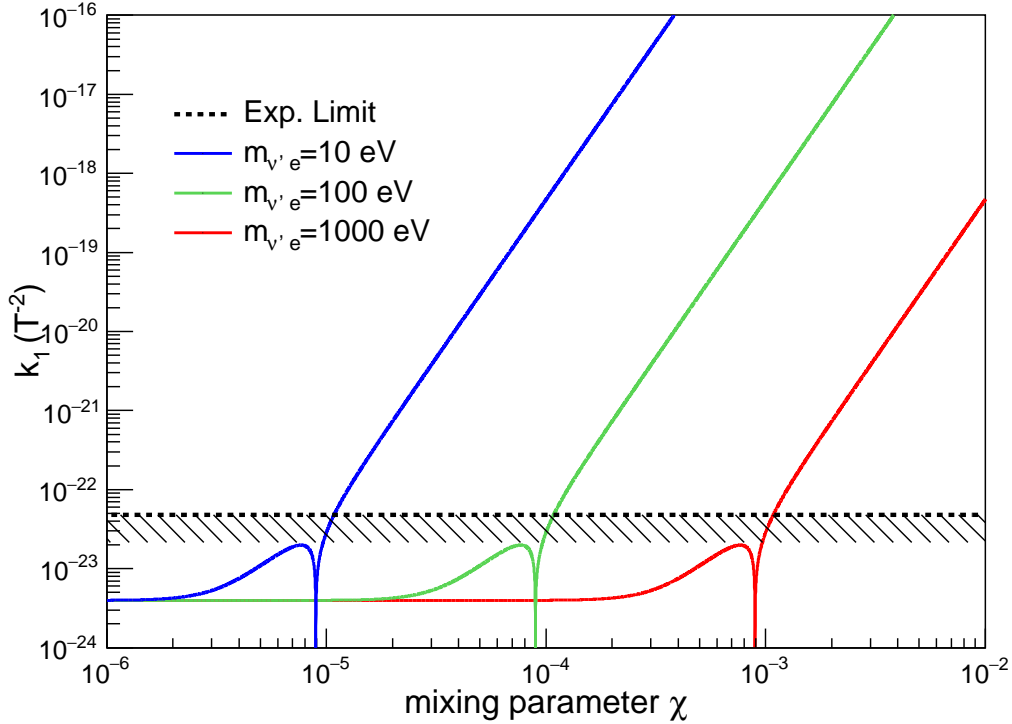}
}
\caption{Magnitude of the ellipticity parameter $k_1$ as a function of mixing parameter $\chi$.
The current experimental limit is also shown in the graph.
The region below the experimental limit is allowed.
The three lines correspond to different masses of the DS neutrino $m_{\nu^{'}_e}$.
The dips in the graph correspond to where $D\simeq 0$, where the difference of refractive indices is about $B^4\times O(a^2, b^2, c^2)$ and the space is almost perfectly isotropic.
The polarization rotation induced in the magnetic field is much smaller than the values shown in the figure (see also Appendix about this point)The area to the left of the dip corresponds to $D>0$, and the area to the right of the dip corresponds to $D<0$.}
\label{fig_kando1}       
\end{center}
\end{figure}

Current experimental sensitivity for $k_1$ is $4.8\times10^{-23}$ (T$^{-2}$), 12 times worse than the QED predicted value.
From this result, we can get a limit on the mixing parameter $\chi$ described in section 3.
Figure \ref{fig_kando1} shows the calculated value of birefringence given in Eqs. (\ref{eq_ellip1})-(\ref{eq_ellip4}) as a function of the mixing parameter $\chi$.
We assume the coupling constant in DS is the same as that in the SM $g'_V=g_V$ and $g'_A=-g'_V$, and use Eq. (\ref{expressionsabc}) to get $a, b,$ and $c$. 
The current experimental limit from the VMB experiment is also shown on the graph.
Dips correspond to where $D\simeq 0$, where the difference of refractive indices is about $B^4\times O(a^2, b^2, c^2)$ and the space is almost perfectly isotropic.
Then the ellipticity induced in the magnetic field is much smaller than the values shown in the figure (see also the Appendix about this point).
From this figure, the experimental limit on the mixing parameter $\chi$ can be given as
\begin{eqnarray}
\chi < 10^{-6} \times (m_{\nu'_e} (\textrm{eV})).
\end{eqnarray}

\subsection{A new experimental scheme to observe parity violation directly}  

The effect of parity violation appears as $\phi$ or $\theta$.
The traditional experimental scheme for measuring VMB, described above, measures the ellipticity as Eqs. (\ref{eq_ellip1})-(\ref{eq_ellip4}) and deduces the magnitudes of $\phi$ and $\theta$.
Since QED itself induces VMB, experiments of this type always have the QED effect as background.
An experimental scheme in which no signal appears within the QED Lagrangian while parity-violating effects appear is desired.
This means that any signal can be attributed specifically to parity-violating effects.
Here, we propose a new experimental scheme to achieve this.

\begin{figure}[]
\begin{center}
\resizebox{\textwidth}{!}{%
  \includegraphics{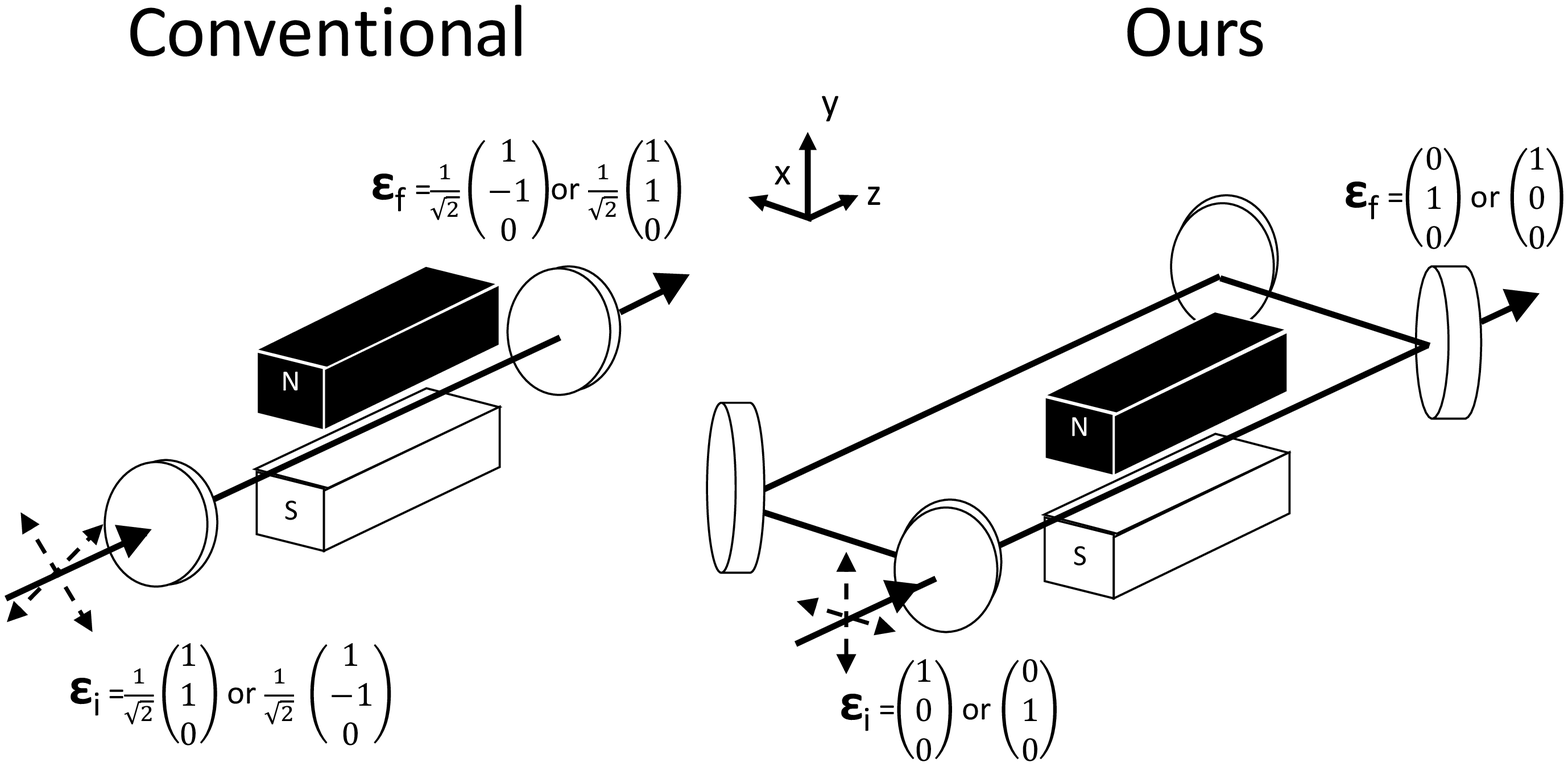}
}
\caption{A schematic of the conventional VMB setup and the new design.
The major difference is the input polarization of the laser and the Fabry-Perot resonator.
In our scheme, the input polarization of the laser is aligned to be parallel or perpendicular to the magnetic field, and the four-mirror Fabry-Perot resonator is used to amplify the parity violation signal.}
\label{fig_drawing}       
\end{center}
\end{figure}

Figure \ref{fig_drawing} shows the new experimental scheme.
$\bm{\epsilon}_{\mathrm{i}}$ is the input polarization state and $\bm{\epsilon}_{\mathrm{f}}$ is the polarization state which is perpendicular to the initial state and does not appear from the QED Lagrangian.
The conventional VMB experimental scheme is also shown in the figure for comparison.
There are two major difference between the traditional scheme and our new design, which are described in the following chapter.

\subsubsection{Polarization of the laser}
The polarization of the injection light, $\bm{\epsilon}_{\mathrm{i}}$, is aligned parallel or perpendicular to the direction of the magnetic field, $\bm{\epsilon}_{\parallel}$ or $\bm{\epsilon}_{\bot}$.
In this configuration, the polarization vector, after going through the magnetic field, can be calculated as

\begin{eqnarray}
\bm{\epsilon}_\parallel &\rightarrow& \left\{\begin{array}{ll}
\left((-i\sin\Psi+\cos2\phi\cos\Psi)\bm{\epsilon}_\parallel +\sin\Psi\sin2\phi\bm{\epsilon}_\bot\right)/\cos2\phi & (D>0)\\\label{eq_original1}
(\cosh\Psi+i\sinh\theta\sinh\Psi)\bm{\epsilon}_\parallel-\cosh\theta\sinh\Psi\bm{\epsilon}_\bot  & (D<0)
\end{array} \right.\nonumber\\
&~&\\
\bm{\epsilon}_\bot &\rightarrow& \left\{\begin{array}{ll}
\left(\sin2\phi\sin\Psi\bm{\epsilon}_\parallel +(i\sin\Psi+\cos2\phi\cos\Psi)\bm{\epsilon}_\bot\right)/\cos2\phi & (D>0)\\\label{eq_original2}
-\cosh\theta\sinh\Psi\bm{\epsilon}_\parallel+(\cosh\Psi-i\sinh\theta\sinh\Psi)\bm{\epsilon}_\bot  & (D<0)
\end{array} \right.\nonumber\\
&~&
\end{eqnarray}

The QED Lagrangian only gives $g_V=-e$ and $g_A=0$; thus $\phi=0$.
The component perpendicular to the initial vector does not appear after going through the magnetic field.
On the other hand, the parity violation effect appears as the real part.
If we can detect the perpendicular component in this situation, that would be direct evidence for parity violation.

One important feature of this configuration is that the perpendicular signal appears as a real number, not as an imaginary number as in the usual VMB configuration (compare Eqs. (\ref{eq_traditional1}), (\ref{eq_traditional2}) with Eqs. (\ref{eq_original1}), (\ref{eq_original2})).
This effect appears in the shape of the output polarization.
Up to order $\Psi$, the output polarization from the new scheme is linear with its polarization axis rotated, while that for the conventional scheme is elliptical with its major axis the same.
The polarization rotation can also be detected with the same order of sensitivity as well as birefringence, which is used and described in detail in \cite{PVLAS}.
Using this configuration, one can measure the magnitude of polarization rotation with a same sensitivity as usual VMB.

\subsubsection{Ring Fabry-P\'erot resonator}
A major method used in a conventional VMB experiment is to use a two mirrors Fabry-P\'erot resonator to enhance the interaction length $L$ about by a factor of $F$.
The finesse of the Fabry-P\'erot resonator is usually more than $10^5$ and this significantly amplifies the magnitude of the signal.
However, this method cannot be applied to enhance the effect of $\Psi$ with the conventional configuration.
Considering that the polarization changes when light is reflected by the mirror, $\bm{\epsilon}_\bot \rightarrow -\bm{\epsilon}_\bot$, the perpendicular component calculated in Eqs. (\ref{eq_original1})-(\ref{eq_original2}) will be canceled out in a round trip.
Thus, in order to enhance $\phi$ and $\theta$, one cannot use a two mirrors Fabry-P\'erot resonator.

The ring Fabry-P\'erot resonator consists of four mirrors.
This type of resonator can accumulate and rotate the light in one direction, clockwise or counterclockwise.
This property ensures that the signs of polarization vectors are maintained in a round trips, $\bm{\epsilon}_\parallel \rightarrow \bm{\epsilon}_\parallel$ and $\bm{\epsilon}_\bot \rightarrow \bm{\epsilon}_\bot$, as depicted in Fig. \ref{fig_drawing}.
This type of Fabry-P\'erot resonator can enhance the magnitude of the signal by about a factor of its finesse $F$.

\subsubsection{Sensitivity of the new setup}
Based on the new setup, we calculate the expected sensitivity.
We define polarization rotation parameter $k_2$ as
\begin{equation}
k_2= (polarization~rotation)\lambda/\pi|B|^2L,
\end{equation}
where the $(polarization~rotation)$ is given from either Eqs. (\ref{eq_original1})-(\ref{eq_original2}).

\begin{figure}[]
\begin{center}
\resizebox{\textwidth}{!}{%
  \includegraphics{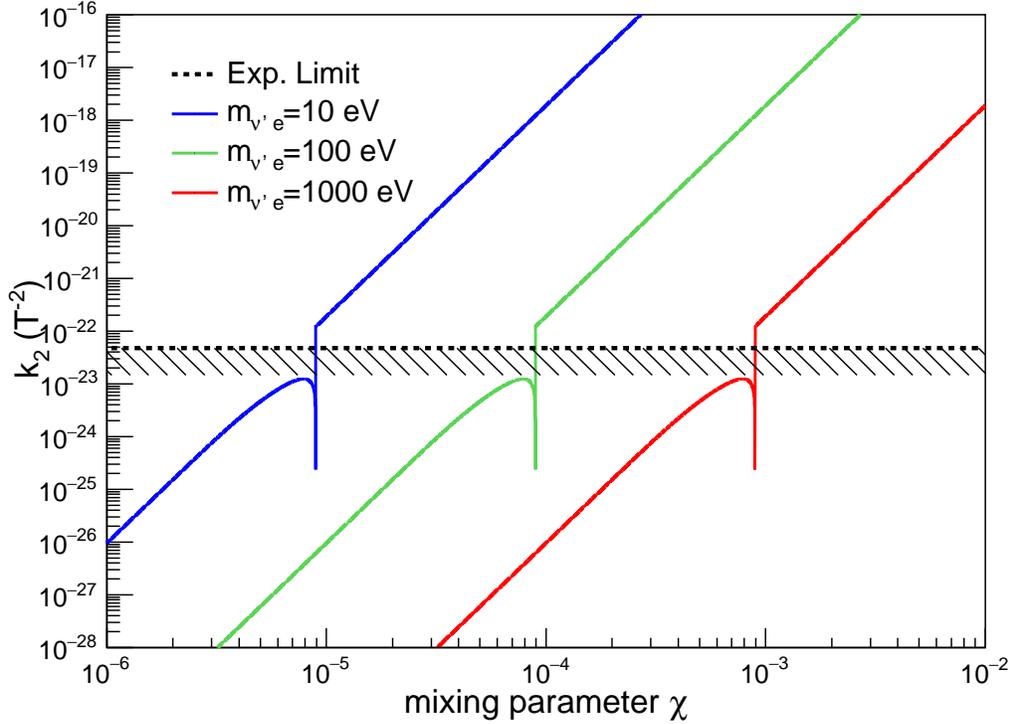}
}
\caption{Magnitude of the polarization rotation parameter $k_2$ as a function of the mixing parameter $\chi$.
The experimental limit expected for polarization rotation is also shown in the graph.
The region below the experimental limit is the allowed region.
The three lines correspond to different masses of the DS neutrino $m_{\nu^{'}_e}$.
Each dip corresponds to $D=0$, where the space is almost perfectly isotropic, and the polarization rotation induced in the magnetic field is much smaller than the values shown in the figure.
The area to the left of the dip corresponds to $D>0$, and the area to the right of the dip corresponds to $D<0$.}
\label{fig_kando2}       
\end{center}
\end{figure}

The current experimental limit on $k_1$ is 12 times worse than the QED predicted value.
From this result, we can obtain a limit on the mixing parameter $\chi$, described in section 3.
Figure \ref{fig_kando2} shows the calculated size of the polarization rotation $k_2$ as a function of the mixing parameter $\chi$.
As in Figure \ref{fig_kando1}, we assume the coupling constant in DS is the same as that in the SM $g'_V=g_V$ and $g'_A=-g'_V$, and use Eq. (\ref{expressionsabc}) to get $a, b,$ and $c$. 
The current experimental limit from the VMB experiment is also shown in the graph.
The experimental limit for polarization rotation is also drawn, assuming that the sensitivity on polarization rotation is as much as that on ellipticity.
The dips in the graph correspond to where $D \simeq 0$, where the difference of refractive indices is about $B^4 \times O(a^2, b^2, c^2)$ and the space is almost perfectly isotropic.
Then the ellipticity induced in the magnetic field is much smaller than the values shown in the figure (see also the Appendix about this point).
One important feature of the figure is that in the $\chi\rightarrow 0$ limit, the magnitude of the polarization rotation $k_2$ also goes to zero.
This means that the new measurement scheme has zero background in the QED regime.
If signal appears in the new scheme, it will be evidence of new physics.

\section{Conclusion}
We examined the possibility of probing effects from dark sector using a vacuum magnetic birefringence experiment.
The effect from the DS can be in general violate parity conservation.
Since the magnitude of birefringence is inversely proportional to the fourth power of mass, the contribution from the dark sector can be as large as the effect from QED.
What is more, since the sensitivity of the VMB experiment does not depend on the mass of the DS photon, it can address the unified DM region, which explain many astrophysical experiments. 
The sensitivity based on the current experimental limit are also estimated, and a new scheme that can directly observe the parity violation effect is proposed.

\section*{Acknowledgements}
The authors thank G.-C. Cho, Toshiaki Kaneko, Kiyoshi Kato, Koichi Seo, N. Jones, and Thomas G. Myers for fruitful discussions and support.

\section*{Appendix}

In this Appendix, we derive the refractive indices in the general case, with vector $g_V$ as well as axial vector $g_A$ couplings.  

The effective action is written as 
\begin{eqnarray}
\mathcal{L}_{\mathrm{eff}}=-\mathcal{F}+ a~ \mathcal{F}^2 +b~ \mathcal{G}^2+ i c~ \mathcal{F}\mathcal{G}, \label{simplified H-E}
\end{eqnarray}
with the constants $(a, b, c)$ given in Eqs.(\ref{a})-(\ref{c}).
When we probe the DS, these coefficients should be multiplied by the forth power of the mixing parameter between the SM and the DS $(\chi)^4$ in Eq. (\ref{applied to DS}), or $(\chi_{\nu'})^4$ in Eq. (\ref{applied to DS neutrino}).

We study the propagation of a laser beam with an angular frequency $\omega$ under a strong magnetic field.
We consider the case where the background magnetic field is weak.
The effective Lagrangian including $\mathcal{G}$ and $\mathcal{F}\mathcal{G}$ terms has already been examined by C. Rizzo {\it et al.}\cite{Rizzo}.
Here, we specify the explicit form of the effective Lagrangian of the vacuum, which includes the parity violation effects coming from the mixing between the real world and the DS.
Then, the real part and the imaginary part of the refraction coefficients can be separately understood, and the optical properties such as birefringence, dichroism, and etc can be elucidated, since the coefficients are known explicitly.
This appendix is to make the paper self-contained.   

The electric and magnetic fields consist of two contributions, one from the laser beam, denoted with the suffix $\gamma$, and the other from the background fields, denoted with the prime.
\begin{eqnarray}
\bm{E}=\bm{E}_{\gamma},~\mathrm{and}~\bm{B}=\bm{B}_{\gamma}+ \bm{B}',
\end{eqnarray}
where $\bm{E}_{\gamma}$ and  $\bm{B}_{\gamma}$ are the fields of the laser beam propagating in the z direction with a wave vector $k$ and a frequency $\nu=\omega/2\pi=v k$, where $v$ is the phase velocity of the laser beam. 
The vector potential, electric field, and magnetic field of the laser beam are given by
\begin{eqnarray}
\bm{A}_{\gamma}(t, z) &=& \bm{\epsilon} e^{-ik(vt-z)}=\bm{\epsilon} A(t, z); \\
\bm{E}_{\gamma}(t, z) &=& -\bm{\epsilon} (-ikv) A(t, z),~\mathrm{and}~\bm{B}_{\gamma}(t, z) = -\tilde{\bm{\epsilon}} (ik) A(t, z), \label{field expansion} 
\end{eqnarray}
where the polarization vector of the laser beam is given by its (0, 1, 2, 3) components as
\begin{eqnarray}
\bm{\epsilon}=(0, \epsilon_{\parallel}, \epsilon_{\bot}, 0), ~\mathrm{and}~\tilde{\bm{\epsilon}}=(0, \epsilon_{\bot}, -\epsilon_{\parallel}, 0).
\end{eqnarray}

The magnetic field $\bm{B}$ is applied in the (x, z) plane, with an angle $\theta_B$ from the propagation direction z of the laser beam.
Therefore, the polarization vector in the x direction is called the parallel $(\parallel)$, and that in the y direction is called the perpendicular $(\bot)$. 
Note that in usual VMB experiments, $\theta_B=\frac{\pi}{2}$.

Substituting the expansion in Eq. (\ref{field expansion}) into Eq. (\ref{simplified H-E}), we have
\begin{eqnarray}
\mathcal{L}_{\mathrm{eff}}=\frac{1}{2} A_{\gamma}(t, z)\pmatrix{\epsilon_{\parallel},~ \epsilon_{\bot}}  \hat{M} (v)\pmatrix{\epsilon_{\parallel} \cr \epsilon_{\bot}} A_{\gamma}(t, z),
\end{eqnarray}
where the velocity dependent 2 $\times$ 2 matrix $\hat{M} (v)$ is given by
\begin{eqnarray}
\hat{M}(v)=\pmatrix{(1-v^2) (1-a|B|^2)-2b v^2 (\sin \theta_B |B|)^2 &,& icv~(\sin \theta_B |B|)^2 \cr  icv~(\sin \theta_B |B|)^2 &,& (1-v^2) (1-a|B|^2)-2a (\sin \theta_B |B|)^2}.
\end{eqnarray}
Now the equation of motion for the laser field reads 
\begin{eqnarray}
\hat{M} (v)\pmatrix{\epsilon_{\parallel} \cr \epsilon_{\bot}}=0,
\end{eqnarray}
To have a non-zero solution for $\bm{\epsilon}$, we have to impose $\det \hat{M}(v)=0$.  This condition gives two solutions for $v^2$.  They are
\begin{eqnarray}
v^2_{\pm}= 1- (\sin \theta_B |B|)^2 \left\{ (a+b) \pm \sqrt{(a-b)^2-c^2} \right\} +O(|B|^4).
\end{eqnarray}
These give the magnitude of the velocity $v$ and the refraction coefficient $n=c/v=1/v$, in the usual case with the small magnetic field $|B|$, as follows:
\begin{eqnarray}
v_{\pm}&=& 1- \frac{1}{2}(\sin \theta_B |B|)^2 \left\{ (a+b) \pm \sqrt{(a-b)^2-c^2} \right\}, \mathrm{and~hence} \\
n_{\pm}&=&1+ \frac{1}{2}(\sin \theta_B |B|)^2 \left\{ (a+b) \pm \sqrt{(a-b)^2-c^2} \right\}. \label{refraction coefficients}
\end{eqnarray}
Note that our calculations ignore $O(|B|^4)$ terms, and $v_{\pm}$ and $n_{\pm}$ seem to degenerate when $D=(a-b)^2-c^2=0$.
However, near $D=0$, the $O(|B|^4)$ terms still exist if we calculate the determinant $det(M(v))$without approximation, so $v_{\pm}$ and $n_{\pm}$ do not degenerate perfectly.
The criterion is is given by
\begin{equation}
|c-(a-b)| \ll B^2 \times O(a^2, b^2, c^2).
\end{equation}
Note that here $a$, $b$, and $c$ are about the same order of magnitude $O(a)\simeq O(b)\simeq O(c)$.
Numerically, since $B^2(a+b)\simeq 10^{-24}\times(B[\mathrm{T}])^2$, this gives
\begin{equation}
|c-(a-b)| \ll 10^{-24}\times(B[\mathrm{T}])^2 \times O(a),
\end{equation}
which is satisfied only in a very small parameter space. 
Thus in the following, we consider only when $|c-(a-b)|$ is larger than $B^2 \times O(a^2)$.
Even when $|c-(a-b)| \ll B^2 \times O(a^2)$, that means $v_{\pm}$ and $n_{\pm}$ almost degenerate, and the magnitudes of ellipticity and polarization rotation of the light are tiny.

If $c\ne0$, we need to be careful on whether the refraction coefficients are real or imaginary.
When the discriminant $D=(a-b)^2-c^2 > 0$, the refraction coefficients are real.
However, when $D=(a-b)^2-c^2 < 0$, the refraction coefficients have imaginary parts.  
In this case, Eq. (\ref{refraction coefficients}) can be written as $n_{\pm}=Re~ n_{\pm} + i~Im~ n_{\pm}$, and the $Re~ n_{\pm}$ and $Im~ n_{\pm}$ are, respectively, the ordinary refraction and the absorption coefficients,
\begin{eqnarray}
&&Re~n_{+}=Re~ n_{-}=1+ \frac{1}{2}(\sin \theta_B |B|)^2 (a+b), \\
&&Im~ n_{+}=-Im ~n_{-}= \frac{1}{2}(\sin \theta_B |B|)^2 \sqrt{c^2-(a-b)^2}.
\end{eqnarray}
Therefore, when $D \ge 0$, the magnetic field induces ``birefringence'', and when $D < 0$, it induces ``dichroism''.

The polarization vector $\bm{\epsilon}_{\pm}$ having the refraction index $n_{\pm}$ is given by 
\begin{eqnarray}
\bm{\epsilon}_{\pm} \propto \left(0, -ic, (a-b) \pm \sqrt{(a-b)^2-c^2}, 0 \right) \propto \left(0, (a-b) \mp \sqrt{(a-b)^2-c^2}, ic, 0 \right)
\end{eqnarray}
for $D>0$, and
\begin{eqnarray}
\bm{\epsilon}_{\pm} \propto \left(0, -ic, (a-b) \pm i\sqrt{c^2-(a-b)^2}, 0 \right) \propto \left(0, (a-b) \mp i\sqrt{(a-b)^2-c^2}, ic, 0 \right)
\end{eqnarray}
for $D<0$.
Note that the two polarizations $\bm{\epsilon}_{\pm}$ are not perpendicular in general, since $\bm{\epsilon}_{+}^{\dagger} \cdot \bm{\epsilon}_{-} \ne 0$.

In the case of $c=0$, the two polarizations remain linear and perpendicular with each other, that is 
\begin{eqnarray}
\bm{\epsilon}_{+}=\bm{\epsilon}_{\bot} ~\mathrm{and}~ n_{+}=n_{\bot}&=&1+ a (\sin \theta_B |B|)^2, ~\mathrm{while} \\
\bm{\epsilon}_{-}=\bm{\epsilon}_{\parallel} ~\mathrm{and}~ n_{-}=n_{\parallel}&=&1+ b (\sin \theta_B |B|)^2.
\end{eqnarray}
This reproduces the ordinary Heisenberg-Euler results in QED.

Using the calculations here, we can discuss generally the vacuum birefringence effects with vector and axial vector couplings.
Note that in general, parity ($\mathcal{P}$) can be violated, which introduces the third term $i c \mathcal{F}\mathcal{G}$ into the effective Lagrangian.  This term preserves the charge conjugation symmetry ($\mathcal{C}$), but violates $\mathcal{P}$, $\mathcal{CP}$ and the time reversal symmetry ($\mathcal{T}$), while the $\mathcal{CPT}$ symmetry is still conserved.

%
%


\begin{thebibliography}{99}

\bibitem{BMV}
A. Cand\`ene et al., Eur. Phys. J. {\bf D 68}, 16 (2014).

\bibitem{PVLAS}
F. Della Valle et al., Eur. Phys. J. {\bf C 76}, 24 (2016).

\bibitem{OVAL}
X. Fan, et al., Eur. Phys. J. {\bf D 71}, 308 (2017); T. Yamazaki et al., NIM {\bf A 833}, 122 (2016); 
X. Fan, Master Thesis, the University of Tokyo, March (2017).

\bibitem{Heisenberg-Euler}
W. Heisenberg and H. Euler, Z. Phys. {\bf 98} (1936) 714 ; W. Heisenberg and H. Euler, arXiv:physics/0605038.

\bibitem{Baier1}
R. Baier and P. Breitenlohner, Act. Phys. Austriaca {\bf 25} (1967) 212.

\bibitem{Baier2}
R. Baier and P. Breitenlohner, Nuov. Cim. B {\bf 47} (1967) 117.

\bibitem{Nuovo}
D. B. Melrose and R. J. Stoneham, Nuovo Cim. A {\bf 32} (1976) 435.

\bibitem{Toll}
J. S. Toll, Ph.D. thesis, Princeton University,1952 (unpublished).

\bibitem{lsw1}
B. D\"obrich, H. Gies, N. Neitz, and F. Karbstein, Phys. Rev. Lett. {\bf 109} (2012) 131802

\bibitem{lsw2}
B. D\"obrich, H. Gies, N. Neitz, and F. Karbstein, Phys. Rev. D {\bf 87} (2013) 025022 

\bibitem{review of H-E}
See also review articles: G. V. Dunne, arXiv:hep-th/0406216; I. Huet, M. R. De Traubenberg, C. Schubert, arXiv:1112.1049;  F. Karbstein, arXiv:1611.09883.

\bibitem{Dittrich}
W. Dittrich and H. Gies, Springer Tracts Mod. Phys. (2000) 166.

\bibitem{Dunne}
G. V. Dunne, From fields to strings, {\bf 1}, (2004) 445, arXiv:hep-th/0406216.

\bibitem{Rizzo1}
R. Battesti and C. Rizzo, Rept. Prog. Phys. {\bf 76} (2013) 016401, arXiv:1211.1933.

\bibitem{Karbstein}
H. Gies and F. Karbstein, JHEP  {\bf 1703} (2017) 108, arxiv:1612.07251.

\bibitem{generalized H-E}
K. Yamashita, X. Fan, S. Kamioka, S. Asai, and A. Sugamoto, Prog. Theor. Exp. Phys. (to be published), arXiv:1707.03308.

\bibitem{Schwinger}
J. Schwinger, Phys. Rev. {\bf 82}, 664 (1951).

\bibitem{Feynman}
R. P. Feynman, Rev. Mod. Phys. {\bf 20}, 367 (1948).

\bibitem{Fock and Nambu}
V. Fock, Physik. Z. Sowjetunion, {\bf 12}, 404 (1937); Y. Nambu, Prog. Theor. Phys. {\bf 5}, 82 (1950).

\bibitem{Holdom}
B. Holdom, Phys. Lett. B {\bf 166}, 196 (1986).

\bibitem{HP 1}
J. Jaeckel, in ``Frascati Physics Series Vol. LVI, Dark Forces at Accererators'' (2012), arXiv:1303.1821.

\bibitem{HP 2}
As a comprehensive review on Dark Sector, see J. Alexander {\it et al.}, ``Dark Sectors 2016 Workshop: Community Report'', arXiv:1608.08632.

\bibitem{Jones}
R. C. Jones, J. of the Optical Society of America, {\bf 38}, 671 (1948).

\bibitem{Rizzo}
B. P. Da Souza, R. Battesti, C. Robilliard, and C. Rizzo, Eur. Phys. J. D {\bf 40}, 445 (2006).

\bibitem{RFootRVolkas}
R. Foot, H. Lew, and R. R. Volkas, Phys, Lett. {\bf 272}, 67 (1991).

\bibitem{RFoot}
R. Foot, Int. J. Mod. Phys. A {\bf29}, 1430013 (2014), arXiv:1401.3965.

\bibitem{universal DM}
Nima Arkani-Hamed, et. al., Phys, Rev. D {\bf 79}, 015004 (2009), arXiv:0810.0713.

\end{thebibliography}
\end{document}